\documentclass[
superscriptaddress,
amsmath,
amssymb,
twocolumn,
prd,
showpacs,
floatfix,
]{revtex4-2}

\usepackage{float}
\usepackage{verbatim}
\usepackage{graphicx}          
\usepackage{bm}                
\usepackage{hyperref}          
\usepackage[dvipsnames]{xcolor}

\begin{document}

\title{De-excitations of highly excited $^{11}$B$^*$ and $^{15}$N$^*$ based on the GEMINI++ code}

 \author{Yujie Niu}
 \email{niuyj@ihep.ac.cn}
 \affiliation{Institute of High Energy Physics, Chinese Academy of Sciences,
 Beijing 100049, China}
 \affiliation{School of Physical Sciences, University of Chinese Academy of Sciences, Beĳing 100049, China}
 \affiliation{Center for High Energy Physics, Henan Academy of Sciences, Zhengzhou 450046, China}
 
 \author{Wan-Lei Guo}%
 \email{guowl@ihep.ac.cn (corresponding author)}
 \affiliation{Institute of High Energy Physics, Chinese Academy of Sciences,
 Beijing 100049, China}
 \affiliation{Center for High Energy Physics, Henan Academy of Sciences, Zhengzhou 450046, China}

 \author{Miao He}%
 \email{hem@ihep.ac.cn }
 \affiliation{Institute of High Energy Physics, Chinese Academy of Sciences,
 Beijing 100049, China}
  \affiliation{Center for High Energy Physics, Henan Academy of Sciences, Zhengzhou 450046, China}

 \author{Jun Su}
\email{sujun3@mail.sysu.edu.cn}
\affiliation{Sino-French Institute of Nuclear Engineering and Technology, Sun Yat-Sen University, Zhuhai, 519000, China}
 
\date{\today}

\begin{abstract}
Nuclear de-excitations associated with neutrino-nucleus interactions and nucleon decays are playing an increasingly significant role in neutrino experiments. We explore the GEMINI++ code and estimate its ability to account for the de-excitation processes of highly excited $^{11}$B$^*$ and $^{15}$N$^*$, which can be created in the liquid scintillator and water Cherenkov detectors respectively. It is found that GEMINI++ can not describe the nuclear experimental data of $^{11}$B$^*$ and $^{15}$N$^*$ well. To improve its performance for de-excitations of light nuclei, we modify GEMINI++ and then develop a code of GEMINI++4$\nu$, which can give the best predictions compared with experimental measurements among some widely used statistical model codes.

\end{abstract}

\maketitle

\section{\label{sec1} Introduction}

Excited nuclei can be created by neutrino-nucleon interactions and nucleon decays in neutrino experiments. Subsequent de-excitation processes will emit gamma rays and other massive particles, such as neutrons, protons, deuterons, tritons, and $\alpha$ particles. Detecting emitted particles is very helpful for neutrino experiments to analyze some physical topics. Water Cherenkov (WC) detectors had  utilized de-excitation gamma rays to measure the nucleon decay modes of $p \rightarrow \bar{\nu} K^+$ \cite{Super-Kamiokande:2014otb,Ejiri:1993rh} and neutron invisible decays \cite{Ejiri:1993rh,SNO:2018ydj,Hagino:2018xnt}. Neutrons can be easily identified in liquid scintillator (LS) experiments based on the delayed signal from neutron capture. LS detectors had applied neutron multiplicity to search for  $p \rightarrow \bar{\nu} K^+$~\cite{JUNO:2022qgr,Hu:2021xjz}, the diffuse supernova neutrino background \cite{JUNO:2022lpc}, the strange axial coupling constant \cite{KamLAND:2022ptk,Abe:2023iwq} and  neutron invisible decays \cite{KamLAND:2005pen,JUNO:2024pur,Kamyshkov:2002wp}. 
Liquid Argon detectors have the capability to detect all particles from nuclear de-excitations \cite{ArgoNeuT:2018tvi}\cite{MicroBooNE:2023sxs}, which can potentially be used to improve the energy resolution of accelerator neutrinos \cite{Ershova:2023dbv}, supernova neutrinos \cite{Gardiner:2020ulp} and other possible physics applications \cite{Castiglioni:2020tsu}. As new-generation neutrino experiments such as JUNO\cite{JUNO}, Hyper-Kamiokande\cite{Hyper-K}, and DUNE\cite{DUNE} will begin collecting data in the near future, nuclear de-excitations will play a more and more significant role.

Accurate de-excitation branching ratios are crucial for neutrino experiments to use the information of de-excitation products. However, there are no universally adopted and quantitatively accurate models to describe the full de-excitation chains~\cite{Mancusi:2010tg}. Some theoretical studies focused on the $\gamma$ emission associated with one~\cite{Ejiri:1993rh} and two \cite{Hagino:2018xnt} nucleon decays in $^{16}$O. Experimental measurements of nuclear de-excitations can provide useful and indicative results \cite{Yosoi:2003jjb, N15,Kobayashi:2006gb, Panin:2016div}. However, it is very challenging to measure the complete branching ratios. Consequently, some statistical model codes are applied to calculate de-excitation processes, including TALYS \cite{TALYS} used in Refs.\cite{JUNO:2022qgr,Hu:2021xjz,JUNO:2022lpc,KamLAND:2022ptk,Abe:2023iwq,Gardiner:2020ulp}, ABLA \cite{ABLA} in Ref.\cite{Ershova:2023dbv,Abe:neutrino2024}, SMOKER \cite{SMOKER} in Refs.\cite{Kamyshkov:2002wp} and CASCADE \cite{CASCADE} in Refs.\cite{Yosoi:2003jjb, N15}. Note that none of these codes can adequately account for the measured de-excitation results of light nuclei $^{11}\text{B}^*$ and $^{15}\text{N}^*$ \cite{Yosoi:2003jjb, N15,Hu:2021xjz,Abe:2023iwq,Abe:neutrino2024}. In addition, the output of widely used TALYS is not in an event-by-event format, making it difficult to be integrated into neutrino generators, such as GENIE \cite{Andreopoulos:2015wxa} 
and NuWro \cite{Golan:2012rfa}. Therefore, current neutrino experiments urgently require a de-excitation code that can well describe experimental data and provide de-excitation results on an event-by-event basis. 

The GEMINI++ code is among the most sophisticated codes for handling the complex-fragment formation in heavy-ion fusion experiments \cite{Mancusi:2010tg,Charity:2010wk,Charity:1988zz,Charity:2008huiyi}. In this work, we shall investigate this code and modify it to account for de-excitations of light nuclei ${\rm ^{11}B^*}$ and ${\rm ^{15}N^*}$. The modified GEMINI++ (GEMINI++4$\nu$) can describe the nuclear experimental data well. This paper is organized as follows. In Sec.~\ref{sec2}, we summarize the experimental measurements and statistical model calculations of highly excited ${\rm ^{11}B^*}$ and ${\rm ^{15}N^*}$ de-excitations so far.  Sec. \ref{sec3} describes briefly the de-excitation formalism employed in GEMINI++, and compare its predictions with experimental data. In Sec.~\ref{sec4}, we modify the GEMINI++ code after examining it and its outputs in detail. Finally, a summary will be given in Sec. \ref{sec5}.

\begin{figure*}
  \centering    
  \includegraphics[width=0.48\textwidth]{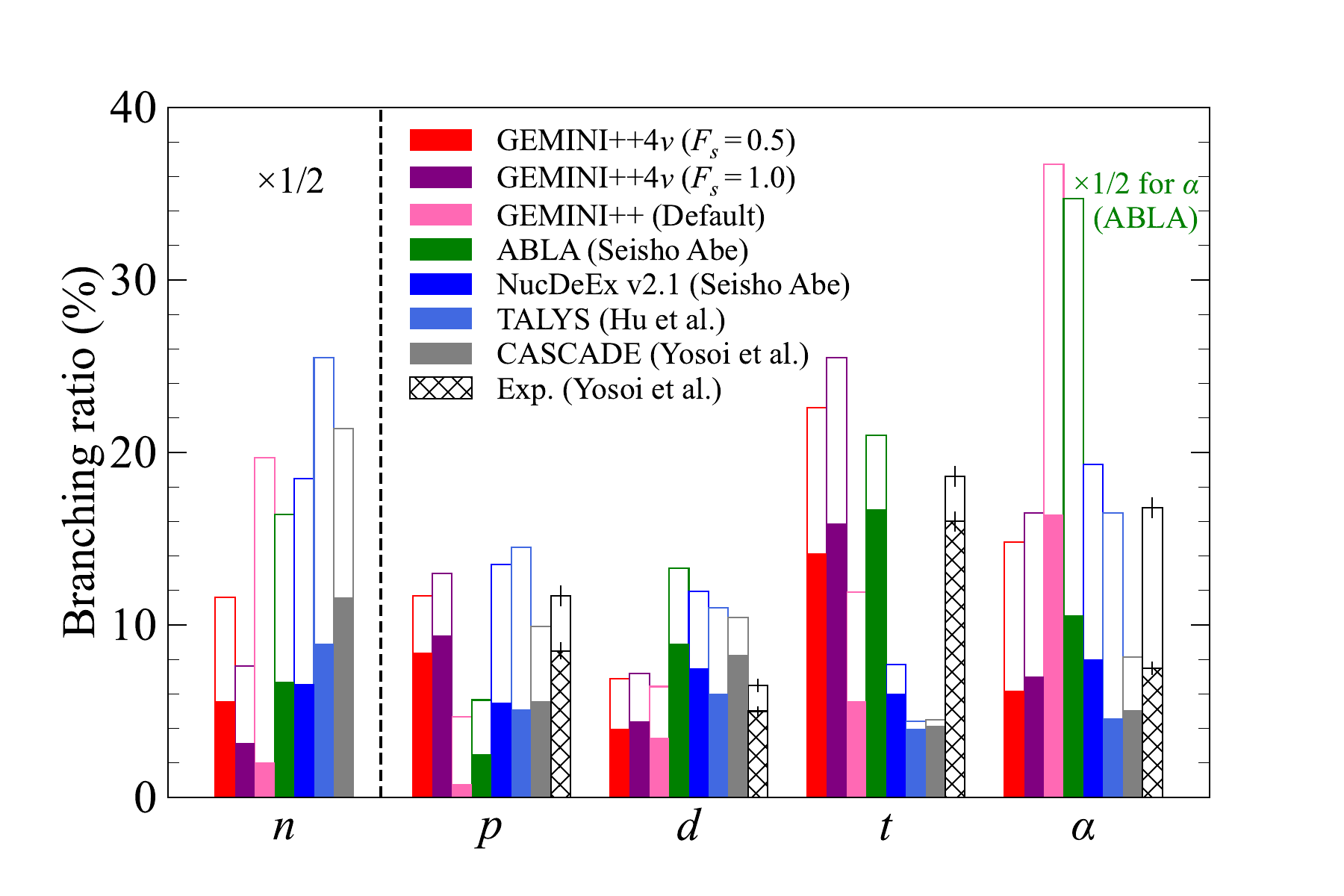}
  \includegraphics[width=0.48\textwidth]{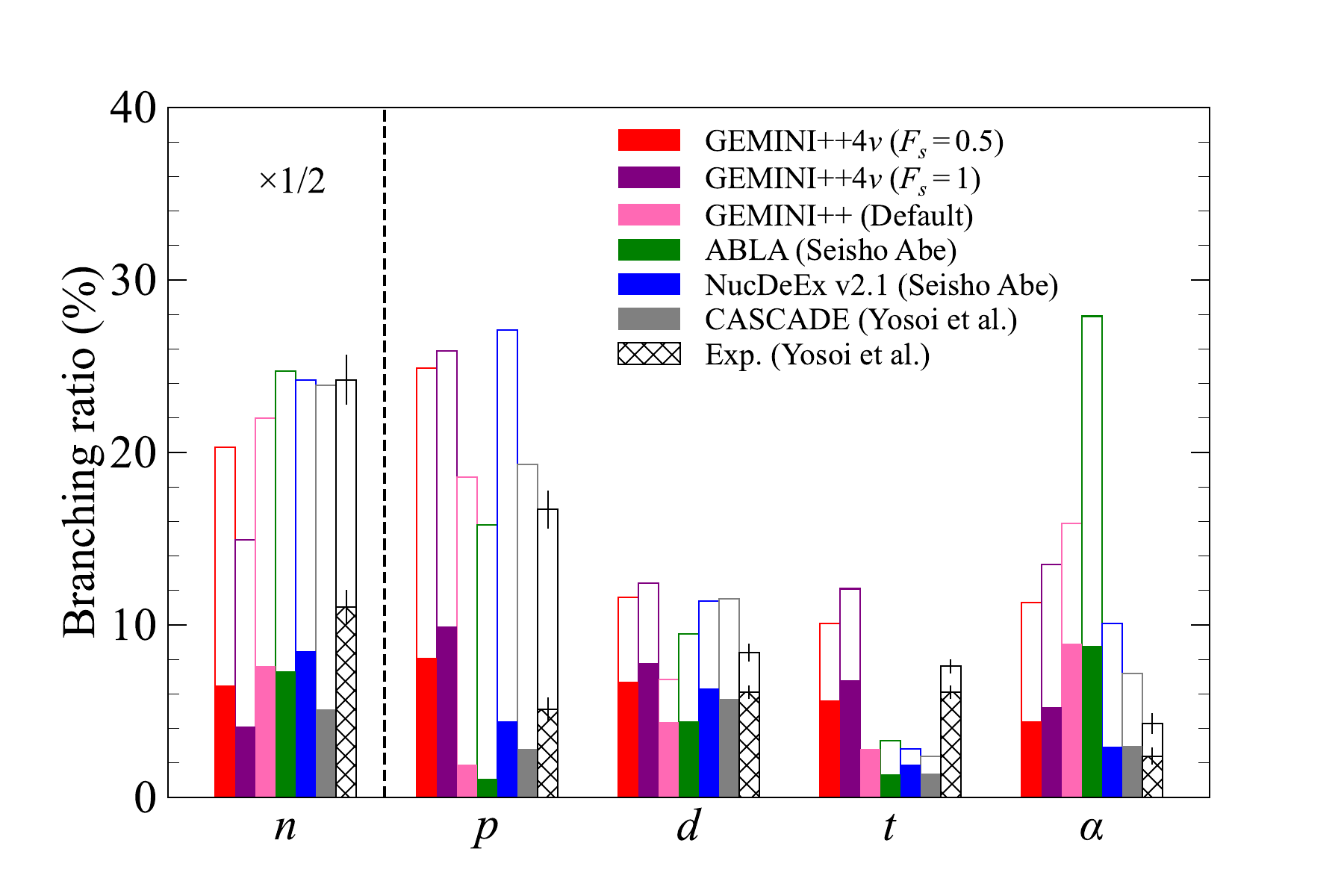}
  \caption{ Comparison of measured and predicted branching ratios of $n, p, d, t, \alpha$ emissions from  de-excitations of $^{11}{\rm B}^*$ (left) with $16 ~{\rm MeV} \leq E_x \leq 35~ {\rm MeV}$ and $^{15}{\rm N}^*$ (right) with $20 ~{\rm MeV} \leq E_x \leq 40 ~{\rm MeV}$. The hatched/colored and blank histograms denote the ‘two-body decay’ and ‘three-body decay’, respectively. }
  \label{fig:Yosoi_Ratio}
\end{figure*}

\section{\label{sec2} De-excitations of highly excited $^{11}$B$^*$ and $^{15}$N$^*$}

In order to assess performances of de-excitation codes, it is necessary to compare their predictions with measurements for interested nuclei. The nuclei with the largest mass fraction in LS and WC detectors are $^{12}$C and $^{16}$O, respectively. Therefore, neutrino-nucleus interactions and nucleon decays can create easily the excited states of $^{11}$B$^*$, $^{11}$C$^*$, $^{15}$N$^*$ and $^{15}$O$^*$ in the case of a nucleon knockout or decay. Subsequent final state interactions will produce lighter excited nuclei. Many experiments have been conducted to study nuclear de-excitations, but only few provide detailed information such as branching ratios of emitting particles and the excitation energy spectrum. It is found that the experimental data from highly excited $^{11}\text{B}^*$ \cite{Yosoi:2003jjb, Panin:2016div} and $^{15}\text{N}^*$ \cite{N15} are available for us to validate the statistical model calculations. In the following parts of this section, we shall firstly introduce the three experimental measurements, and then briefly discuss performances of different de-excitation codes.

Yosoi et al.\cite{Yosoi:2003jjb,N15} measured charged particle emissions from the $s$-hole states of $^{11}{\rm B}^*$ in coincidence with the quasifree $^{12}{\rm C}(p, 2p)^{11}{\rm B}^*$ reaction using a 392 MeV proton beam, where the charged particles include protons, deuterons, tritons and $\alpha$ particles. The detection thresholds are 3.1, 4.0, 4.6 and 4.5 MeV for $p, d, t$ and $\alpha$ emissions, respectively. Note that $^{3}{\rm He}$ events have been included in the $\alpha$ portion. Due to the detector limitation, the residual daughter nucleus after a full de-excitation cascade can not be detected. In fact, the excitation energy $E_x$(res) of residual daughter nuclei can be induced from the excitation energy $E_x$ of $^{11}{\rm B}^*$ and the kinetic energy of a detected charged particle as shown in Fig. 2 of Ref.~\cite{Yosoi:2003jjb}. Then the authors define the ‘two-body decay’ and ‘three-body decay’ in terms of $E_x$(res).
The ‘two-body decay’ region for each particle is defined as $E_x$(res)~$\leq$~Max(5 MeV, $E_{\rm th}$(3-body)), where $E_{\rm th}$(3-body) denotes the threshold energy of massive particle decay in the residual nucleus. The ‘three-body decay’ mainly comes from contributions of the realistic three-body and sequential binary de-excitation processes. Note that ‘two-body decay’ and ‘three-body decay’ are enclosed in single quotes to match the measurements in Refs.~\cite{Yosoi:2003jjb,N15}.
The measured branching ratios of ‘two-body decay’ and ‘three-body decay’ for $E_x =$16-35~MeV have been plotted in the left panel of Fig.~\ref{fig:Yosoi_Ratio}. 
The hatched/colored and blank areas indicate the ‘two-body decay’ and ‘three-body decay’ contributions, respectively.
It is clear that the $t$ and $\alpha$ decays contribute the most to the charged particle emissions from the $s$-hole states of excited $^{11}{\rm B}^*$. Most of $p$, $d$ and $t$ particles belong to the ‘two-body decay’.    

Yosoi et al.\cite{N15} also measured charged particle emissions from de-excitations of $^{15}{\rm N}^*$ produced by the quasifree $^{16}{\rm O}(p, 2p)^{15}{\rm N}^*$ reaction using the same proton beam and detector. The right panel of Fig.~\ref{fig:Yosoi_Ratio} shows the measured branching ratios of charged particles and neutrons from $^{15}{\rm N}^*$ de-excitations for $E_x =$20-40~MeV. The fraction of $t$ and $\alpha$ decays is smaller than that of $p$ and $d$ decays, which is completely different from the $^{11}{\rm B}^*$ case. It is worthwhile to stress that this experiment measured separately the neutron emission of highly excited $^{15}{\rm N}^*$ in the second beam time \cite{N15}. To detect neutrons, this measurement used a small scattering chamber and a neutron multi-detector array system. The lower limit of neutron detection energy is 3.2 MeV \cite{thesis}.
Definitions of ‘two-body decay’ and ‘three-body decay’ for $^{15}{\rm N}^*$ are similar with $^{11}{\rm B}^*$ but the maximum value is chosen from 8 MeV and the corresponding $E_{\rm th}$(3-body). As shown in the right panel of Fig.~\ref{fig:Yosoi_Ratio}, the measured branching ratio of neutrons is about 50\%, which is larger than the contribution of all charged particles. For the neutron emission, the contributions of both the ‘two-body decay’ and ‘three-body decay’ are roughly equal.

\begin{figure}
  \centering
  \includegraphics[width=0.450\textwidth]{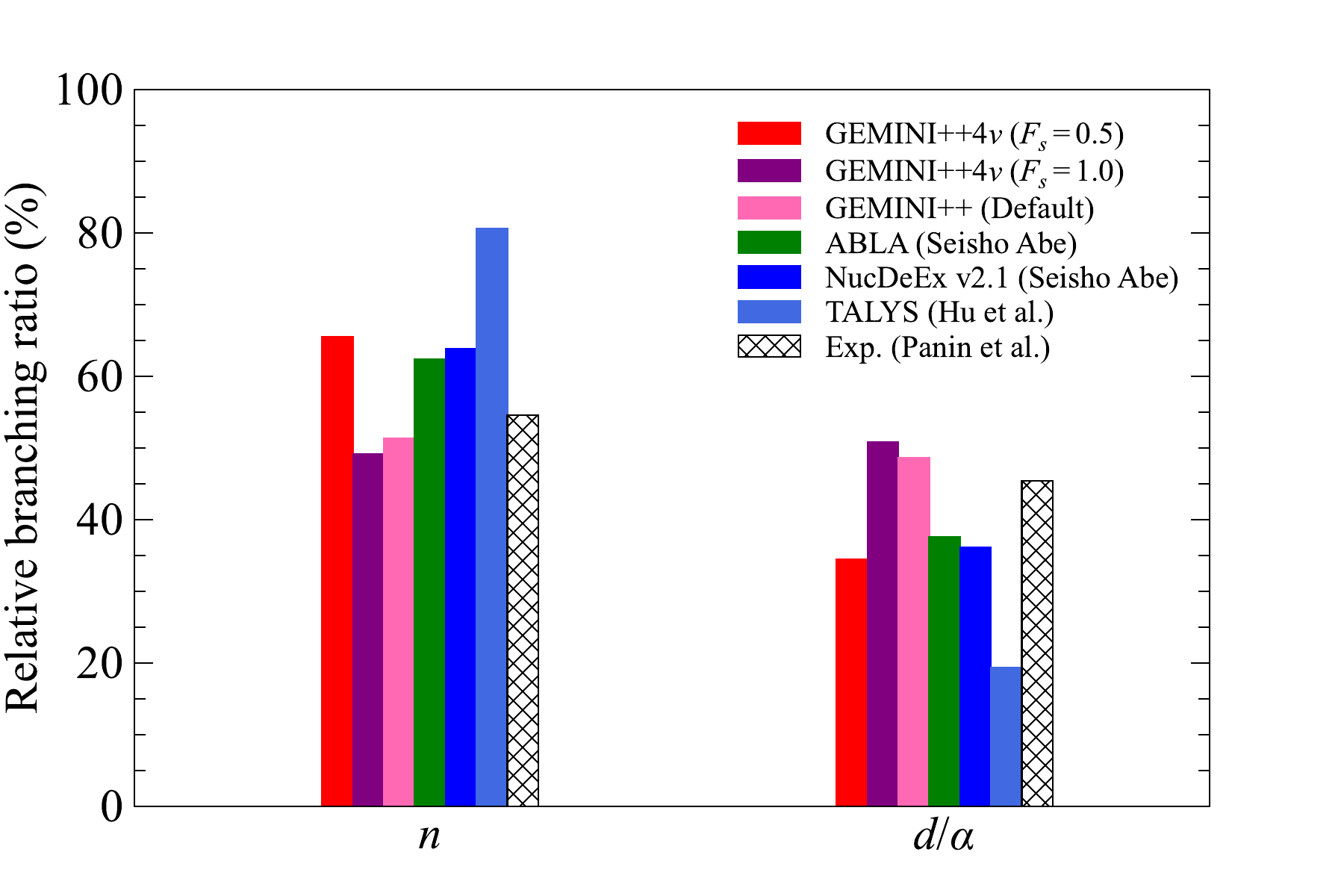}
  \caption{Comparison of measured and predicted relative branching ratios of  $n$ and $d/\alpha$ from $^{11}{\rm B}^*$ de-excitations with $16 ~{\rm MeV} \leq E_x \leq 35~ {\rm MeV}$.}
  \label{fig:Panin_Ratio}
\end{figure}

Panin et al.\cite{Panin:2016div} measured the quasi-free reaction of $^{12}{\rm C}(p, 2p)^{11}{\rm B}^*$ using a $^{12}{\rm C}$  beam at an energy of $\sim$~400~MeV/u as a benchmark. The two outgoing protons, emitting particles and moving residue nuclei of $^{11}{\rm B}^*$ de-excitations are measured for the first time exclusively in complete and inverse kinematics. Benefiting from the subtle design, this experiment can measure the kinematics of the residual nuclei and determine whether the de-excitation process is a two-body decay channel. There is a limitation that only three two-body decay channels are analyzed, which are $^{11}{\rm B}^* \rightarrow n + ^{10}{\rm B}$,  $^{11}{\rm B}^* \rightarrow d + ^{9}{\rm Be}$ and $^{11}{\rm B}^* \rightarrow \alpha$+$^{7}{\rm Li}$. Fig.~4 of Ref.\cite{Panin:2016div} presents the measured distributions of $^{11}{\rm B}^* \rightarrow n + ^{10}{\rm B}$ and all three two-body channels as a function of the excitation energy $E_x$ of $^{11}{\rm B}^*$. Therefore two relative branching ratios of the $n$ and $d/\alpha$ emissions among all three two-body decay modes can be calculated for $E_x =$16-35~MeV as shown in Fig.~\ref{fig:Panin_Ratio}. The $d/\alpha$ ratio includes the contribution from both $^{11}{\rm B}^* \rightarrow d + ^{9}{\rm Be}$ and $^{11}{\rm B}^* \rightarrow \alpha$+$^{7}{\rm Li}$. It is clear that the neutron emission accounts for half of the total emissions among the three two-body decay channels mentioned above.    

Some statistical model codes have been used to predict the de-excitations of $^{11}$B$^*$ and $^{15}$N$^*$ as shown in Fig.~\ref{fig:Yosoi_Ratio}. Yosoi et al. utilized the CASCADE code \cite{CASCADE} to interpret their experimental data. However, this code does not provide a good agreement \cite{Yosoi:2003jjb, N15}, especially for the $t$ and $\alpha$ emissions. Hu et al. used the TALYS code (version 1.95) \cite{TALYS} to model $^{11}$B$^*$ de-excitations \cite{Hu:2021xjz}. The TALYS calculation can partly account for the experimental data as shown in Figs.\ref{fig:Yosoi_Ratio} and~\ref{fig:Panin_Ratio}. Seisho Abe combined the TALYS code (version 1.96) choosing a different nuclear model with Geant4 \cite{GEANT4} to develop an event-by-event simulator of NucDeEx \cite{Abe:2023iwq}. It can give a better result but the branching ratio of tritons is still not reproduced. Recently, Seisho Abe considers the ABLA code \cite{ABLA} to explain the experimental data of $^{11}$B$^*$ and $^{15}$N$^*$. As shown in Fig.\ref{fig:Yosoi_Ratio}, the ABLA predictions are worse than that from the CASCADE and TALYS codes, especially for the $\alpha$ emission \cite{Abe:neutrino2024}. From what has been discussed above, the current statistical model calculations can not account for the measured data well. The large discrepancy in the $t$ emission from both $^{11}$B$^*$ and $^{15}$N$^*$ de-excitations is still visible.

\section{\label{sec3}  GEMINI++ predictions }

GEMINI++, a Monte Carlo code \cite{Mancusi:2010tg,Charity:2010wk,Charity:2008huiyi}, is an improved C++ version based on GEMINI \cite{Charity:1988zz} and designed for handling complex fragment formation in heavy-ion fusion reactions. This code has been extensively used in nuclear physics and got cheerful achievements. In GEMINI++, statistical sequential binary decays of compound nucleus are carried out to perform de-excitation processes, which terminates when either the energy conservation is no longer possible or the process loses in competition with $\gamma$ decay. GEMINI++ provides not only light particle evaporation modes but also fission, complex fragment decay, and gamma emission modes. The partial decay width of a specific evaporation particle is calculated by the Hauser-Feshbach\cite{Hauser:1952zz} or Weisskopf-Ewing (WE) \cite{Weisskopf:1940lhq} formalism. For the focused $^{11}$B$^*$ and $^{15}$N$^*$, the highly excited light nuclei with low angular momentum, the latter formalism is adopted in the GEMINI++ calculation.

In the Weisskopf-Ewing formalism, the partial decay width of a compound nucleus with the excitation energy $E_x$ for the evaporation of particle $i$ is given by \cite{Charity:2008huiyi}
\begin{align} \label{eq1}
\Gamma^{WE}_i = \frac{2S_i+1}{2\pi\rho^0(E_x)} \int \sum_{l=0}^\infty (2l+1) T_l(\varepsilon) \rho(U) d\varepsilon ,
\end{align}
where $S_i$ and $l$ are the spin and orbital angular momenta of the evaporated particle $i$, respectively. Evaporation modes include the emissions of $n$, $p$, $d$, $t$, $^{3} {\rm He}$, $\alpha$, and so on. $T_l(\varepsilon)$ is the transmission coefficient or barrier penetration factor, and depends on the kinetic energy $\varepsilon$ of the evaporated particle.
To ensure full absorption, the incoming-wave boundary-condition model (IWBC)~\cite{Rawitscher:1966iwbc} is employed to calculate $T_l$ with the help of global optical-model potentials \cite{Charity:2008huiyi}.
$U$ in Eq.~(\ref{eq1}) represents $E_x - B_i - E_{rot} - \varepsilon$, where $B_i$ is the separation energy and $E_{rot}$ is the rotational energy of the ground-state configuration \cite{Su:2018veq}. $\rho^0$ and $\rho$ are the spin-independent level densities of the parent and daughter nuclei, and are often approximated by the Fermi-gas form \cite{Charity:2010wk}:
\begin{align} \label{eq2}
\rho(U) \propto \frac{\exp [2 \sqrt{a (U-E_1)}]}{a^{1/4} (U-E_1)^{5/4}},
\end{align}
where $a$ is the level density parameter. The back-shifted term $E_1$ is related with the pairing correction and the shell correction. Note that GEMINI++ introduces the suppression factors $F_s$ to adjust its predictions:
\begin{align} \label{eq3}
\Gamma_{i} = \Gamma^{WE}_i \times F_s.
\end{align}
After the normalization, GEMINI++ gives the final de-excitation results based on $\Gamma_{i}$. In fact, $\Gamma_{i}$ decides the practical evaporation probability for each particle.

There are four parameters used to specify the parent compound nucleus in GEMINI++, the charge number $Z$, mass number $A$, spin $S$ and excitation energy $E_x$. Generally, every nucleus has two types of levels, namely discrete levels and continuum levels. The boundary between discrete and continuum levels is ambiguous. GEMINI++ does not provide the method to deal with discrete states due to the large nuclear data and uncertainties, let along the boundary problem. In the continuum region, the level densities of positive and negative parities are assumed identical, so the decay widths of positive and negative parity states are thus, by assumption, the same and therefore there is no need to specify the parity in the GEMINI++ code.

As mentioned before, the GEMINI++ predictions should be compared with the experimental data. In Refs.~\cite{Yosoi:2003jjb,N15,Panin:2016div}, the highly excited $^{11}$B$^*$ and $^{15}$N$^*$ are produced by an $s$-shell proton knockout from $^{12}$C and $^{16}$O, respectively. Therefore, the spins of both $^{11}$B$^*$ and $^{15}$N$^*$ are 1/2 based on the simplest shell model. In order to maintain consistency with the experimental setups, we input the measured excitation energy spectra of $^{11}$B$^*$ in Ref.~\cite{Yosoi:2003jjb} and $^{15}$N$^*$ in Ref.~\cite{N15} for GEMINI++ to carry this comparison. For the ‘two-body decay’ and ‘three-body decay’ separation, the same selection criteria in Refs.~\cite{Yosoi:2003jjb,N15} have been applied to classify the output events. Meanwhile, we have considered that the emitted particles have different detection thresholds.
The GEMINI++  results have been also plotted in Figs.~\ref{fig:Yosoi_Ratio} and \ref{fig:Panin_Ratio}. It is clear that the GEMINI++ code can not account for the $^{11}$B$^*$ or $^{15}$N$^*$ data well. Only the predicted relative branching ratio of $n$ or $d/\alpha$ from $^{11}{\rm B}^*$ de-excitations approaches to the measured one as shown in Fig.~\ref{fig:Panin_Ratio}. One can easily find that the predicted branching ratios of the $\alpha$ emission are far larger than the measured values in Fig.~\ref{fig:Yosoi_Ratio}. In addition, there is a significant bias for the $p$ emission from $^{11}$B$^*$ and the $t$ emission from $^{15}$N$^*$.

\section{\label{sec4} GEMINI++4$\nu$ predictions }

As discussed above, GEMINI++ can not provide the consistent predictions with the experimental data in Refs.~\cite{Yosoi:2003jjb,N15,Panin:2016div}. It is known that this code is designed for handling complex fragment formation in heavy-ion fusion reactions. Therefore, some parameters and/or settings may be not proper for the de-excitaions of light nuclei of $^{11}$B$^*$ and $^{15}$N$^*$. After studying the GEMINI++ code and examining its outputs in detail, we find that three issues should be addressed. They can significantly change the GEMINI++ predictions. To address the three issues, we have modified the GEMINI++ code. The new code is named as GEMINI++4$\nu$, which is available at the website \cite{G++4v}. In the following parts of this section, we firstly discuss the three issues and describe the corresponding modifications implemented in GEMINI++4$\nu$. Then we present the GEMINI++4$\nu$ results, which can give the best description for the current experiment data.

\subsection{\label{sec4.1} Modifying GEMINI++ }

As shown in Eq.~(\ref{eq2}), the effect of the back-shifted term $E_1$ is equivalent to changing the excitation energy $E_x$ of the compound nucleus. It can significantly change the emissions of massive particles compared with the $\gamma$ emission. Therefore, massive particles will necessitate a higher excitation energy to initiate their evaporation process. To determine this critical excitation energy $E_c$ for $^{11}$B$^*$ de-excitations in the GEMINI++ code, we simulate 5000~events for every interval of 0.1 MeV in the range of $0 ~{\rm MeV} \leq E_x \leq 50~ {\rm MeV}$. The critical energies for 
6 two-body decay modes have been listed in Table~\ref{modes}. 
The predicted $E_c$ for decay modes of $p +^{10}{\rm Be}$ and $t+^{8}{\rm Be}$ exhibit obvious deviations from their separation energies. There should not be such large differences though additional energy is needed to overcome the nuclear effects. This implies that the back-shifted term $E_1$ is not used properly for the case of light nuclei. Compared with the fitting results in Ref.~\cite{vonEgidy:2005rce}, $E_1$ used in GEMINI++ indicates a significant overestimation due to the large  pairing correction calculation. 
Unfortunately, the available $E_1$ is absent for $^{11}$B$^*$, $^{15}$N$^*$ and their daughter nuclei after de-excitations.
Therefore, we modify GEMINI++ through removing the back-shifted term $E_1$ in Eq.~(\ref{eq2}). In this case, the critical energy $E_c$ is closest to the separation energy.
To validate the rationality of the removal of $E_1$, we perform a simple $\chi^2$ calculation based on the formula in Sec.~II 2B of Ref.~\cite{vonEgidy:2005rce} with the term containing $D_{\rm res}$ removed. It is found that most of $^{11}$B$^*$, $^{15}$N$^*$ and their daughter nuclei can give a better agreement with the measured discrete levels due to removal of $E_1$ as discussed in the supplementary material. A similar $\chi^2$ calculation was initially performed to determine the boundary between discrete levels and continuum levels, which gives a rough result of 6 MeV. 
As a result, removing $E_1$ can enhance the $p$ and $t$ branching ratios, and correspondingly reduce the $d$ and $\alpha$ emissions. For $^{11}$B$^*$, this modification will bring the predictions closer to the experimental observations. However, the predicted branching ratios of $p$, $d$ and $n$ from $^{15}$N$^*$ de-excitations will deviate from their measured values.   

\begin{table}
\centering
\caption{ Six $^{11}$B$^*$ de-excitation modes and the corresponding separation energy from theoretical calculations, the critical excitation energy predicted by GEMINI++ and GEMINI++4$\nu$. All numbers are in units of MeV.}
\begin{tabular}{c c c c} \hline  \hline
Modes           & \;\;\;\;\;\ Theory \;\;\;\;\;          & \; GEMINI++\;     & GEMINI++4$\nu$\\  \hline
$n + {^{10}{\rm B}}$              &       11.5      & 12.2      &12.2\\ 
$p + {^{10}{\rm Be}}$             &       11.2      & 17.1      &12.2\\ 
$d + {^{9}{\rm Be}}$              &       15.8      & 16.6      &16.6\\ 
$t + {^{8}{\rm Be}}$              &       11.2      & 20.5      &12.1\\ 
${^{3}{\rm He}}+{^{8}{\rm Li}}$  &       27.2       &  28.6     & 28.6\\ 
$\alpha + {^{7}{\rm Li}}$         &       8.7       & 10.1      &10.1\\ 
\hline 
\end{tabular}
\label{modes}
\end{table}

The daughter nuclei of de-excitations are usually left in discrete levels when they have a relatively low excitation energy. However, this situation does not appear in GEMINI++ since it only considers continuum levels. Here we calculate the influence of discrete levels, which can be implemented through modifying the GEMINI++ outputs on an event-by-event basis. 
Based on the estimated boundary of 6 MeV between discrete and continuum states, we only consider the discrete levels with the excitation energy less than about 6 MeV, and require that the decay information of every added discrete state is known clearly in the NNDC database~\cite{NNDC}.  Modifications of the GEMINI++ output events can be categorized into three types of situations 
for the excited daughter nuclei of $^{11}$B$^*$ and $^{15}$N$^*$:
\begin{itemize}
\item If the excited energy of 
the decay nucleus in the last binary decay is higher than the highest discrete level considered obviously, this GEMINI++ event will be kept;
\item If it is lower than all of the discrete levels, this state will be set as the ground state and the kinematics of the previous decay will be recalculated;
\item  If it lies between two discrete levels, we shall reset this state to the lower level and recalculate the kinematics of the previous decay.
\end{itemize}
Based on the above strategy, we modify the GEMINI++ outputs and find that including discrete levels can enhance the predicted branching ratios by about less than 20\% for both `two-body decay` and `three-body decay' in Fig.~\ref{fig:Yosoi_Ratio}. This is because the aforementioned strategy can increase the kinetic energy of some of the finally emitted particles from below the detection threshold to above it.
Note that the setting of discrete levels does not change the predicted relative  branching ratios in Fig.~\ref{fig:Panin_Ratio}.

\begin{table}
\centering
\caption{ Different settings for suppression factors used in GEMINI++ and GEMINI++4$\nu$ codes.}
\begin{tabular}{l |c |c |c |c |c |c} \hline  \hline
Settings   & \;\; $n$ \;\;   &  \;\; $p$ \;\;   &  \;\; $d$ \;\;    &  \;\; $t$ \;\; &  \, $^{3}$He \,    & \;\; $\alpha$ \;\;  \\ \hline
Default      & 1.0 & 1.0 & 0.5 & 0.5 & 0.5 & 1.0 \\ 
$F_s =1.0$   & 1.0 & 1.0 & 1.0 & 1.0 & 1.0 & 1.0 \\ 
$F_s =0.5$   & 1.0 & 0.5 & 0.5 & 0.5 & 0.5 & 0.5 \\ 
\hline 
\end{tabular}
\label{SF}
\end{table}

As mentioned in Eq.~(\ref{eq3}), the suppression factor $F_s$ plays a tuning role in determining the particle evaporation probabilities. The default settings about $F_s$ have been listed in Table~\ref{SF}, which originates from the de-excitations of heavy nuclei. The suppression factor of 0.5 chosen for $d$, $t$ and $^3$He can significantly reduce their branching ratios. For light nuclei $^{11}$B$^*$ and $^{15}$N$^*$, no available settings for suppression factors can be found. Therefore, we firstly take all suppression factors to be 1.0, namely $F_s=1.0$, which means that the suppression factors actually have no effect. In addition, we set the same $F_s$ for $^{6-7}$Li as for other charged particles, since they take part in the $^{15}$N$^*$ de-excitations. It will be found that the $F_s=1.0$ setting accounts well for the $^{11}$B$^*$ data, but can not reproduce the $^{15}$N$^*$ results in the following subsection. To simultaneously elucidate both the $^{11}$B$^*$ and $^{15}$N$^*$ measurements shown in Fig.~\ref{fig:Yosoi_Ratio}, $F_s=0.5$ for all charged particles has also been assumed as the recommended setting in this work.

\subsection{\label{sec4.2} GEMINI++4$\nu$ results }

Here we discuss the GEMINI++4$\nu$ performance compared with the experimental data in Refs.~\cite{Yosoi:2003jjb, N15, Panin:2016div}. Both $F_s=1.0$ and $F_s=0.5$ scenarios have removed the back-shifted term and added the discrete levels. 
The effects of step-by-step modifications can be found in the supplementary material.
For the $F_s=1.0$ setting, the predicted branching ratios of $^{11}$B$^*$ de-excitations are in a good agreement with the measurements as shown in the left panel of Fig.~\ref{fig:Yosoi_Ratio}. This is the first time that a statistical model code can simultaneously explain all four emissions of charged particles for both two-body and three-body decays of $^{11}$B$^*$, although there are still some minor differences. However, GEMINI++4$\nu$ with $F_s=1.0$ can not account for the $^{15}$N$^*$ de-excitations well. It is clear that obvious differences in the right panel of Fig.~\ref{fig:Yosoi_Ratio} can be found for the $n$, $p$ and $\alpha$ decays. For the relative branching ratios in Fig.~\ref{fig:Panin_Ratio}, GEMINI++4$\nu$ with $F_s=1.0$ gives the better predictions than those from TALYS \cite{Hu:2021xjz}, NucDeEx \cite{Abe:2023iwq} and ABLA \cite{Abe:neutrino2024}.

For the $F_s=0.5$ case, only the suppression factors of $p$ and $\alpha$ are changed from the default value of 1.0 to 0.5. As shown in Fig.~\ref{fig:Yosoi_Ratio}, GEMINI++4$\nu$ with $F_s=0.5$ gives the better agreement with the $^{11}$B$^*$ data than the $F_s=1.0$ setting. Note that the differences of total branching ratios between the measurements and predictions are less than 1\% for the $p$ and $d$ emissions. Meanwhile, the $F_s=0.5$ setting can partially account for the $^{15}$N$^*$ data, especially the $n$ decay included. We believe that smaller suppression factors than 0.5 would better describe the measurements if only the $^{15}$N$^*$ data are compared. In Fig.~\ref{fig:Panin_Ratio}, another comparison shows that GEMINI++4$\nu$ with $F_s=0.5$ can also explain the measured relative branching ratios of $n$ and $d/\alpha$.

\begin{figure*}
  \centering    
  \includegraphics[width=0.45\textwidth]{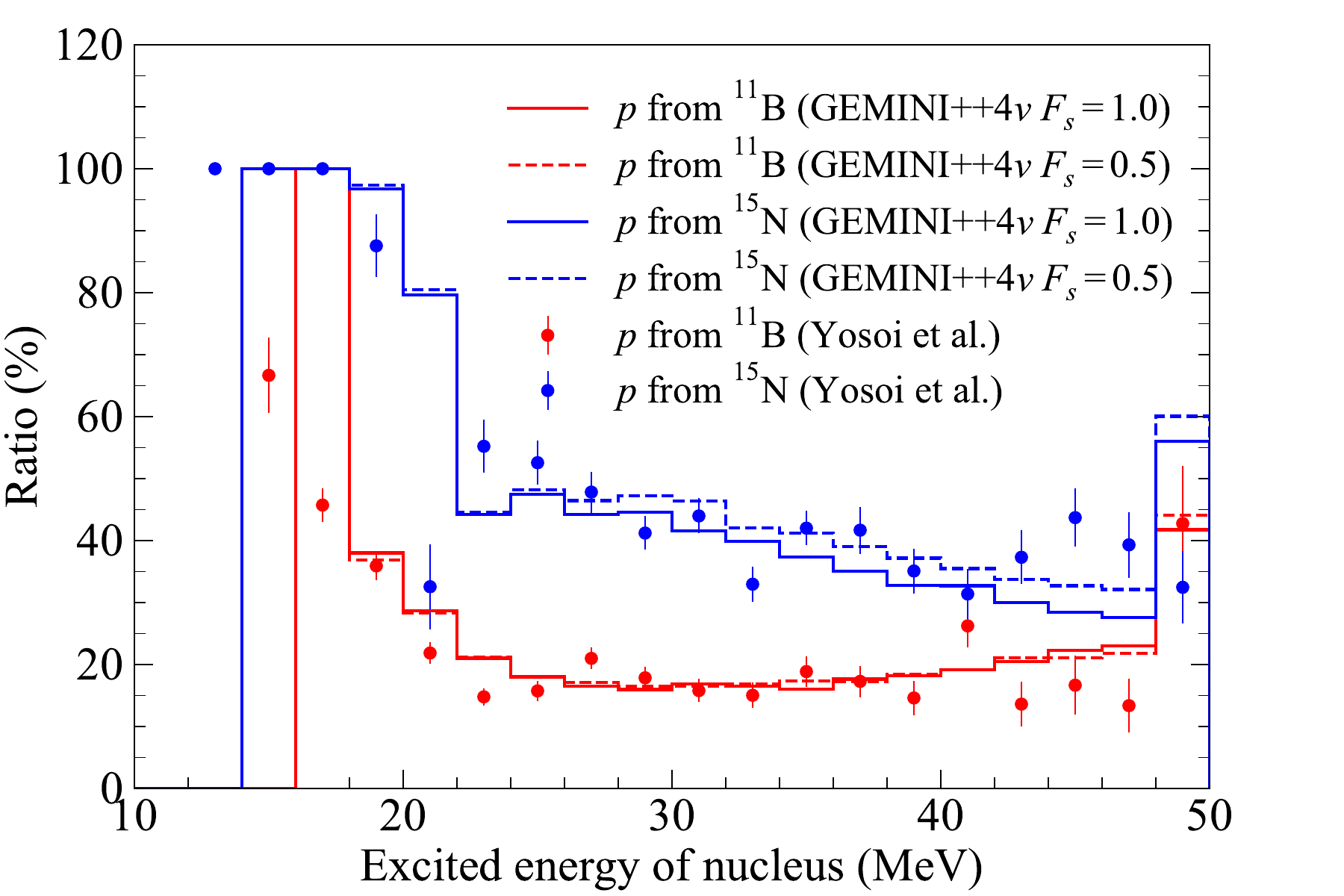}
  \includegraphics[width=0.45\textwidth]{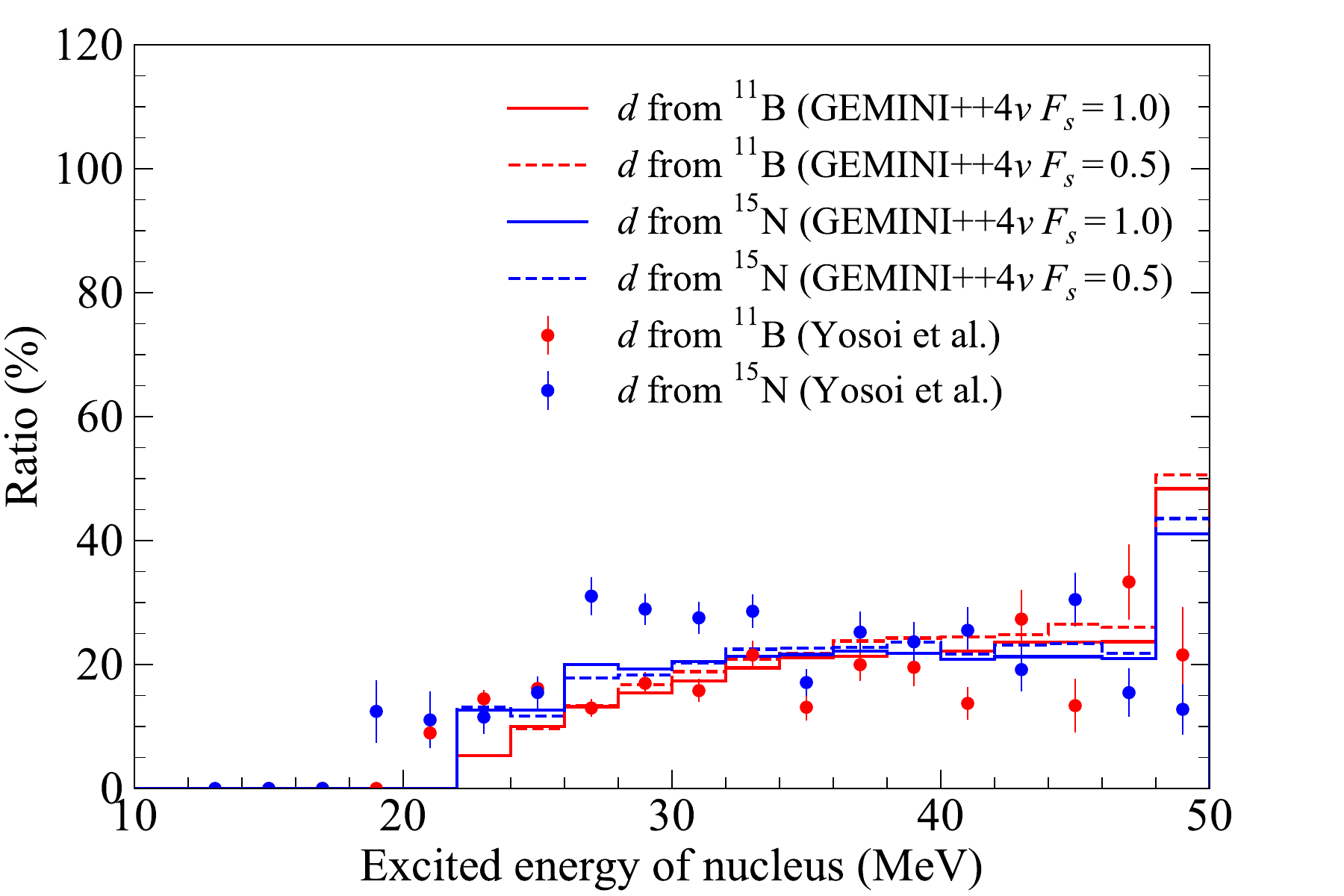}
  \includegraphics[width=0.45\textwidth]{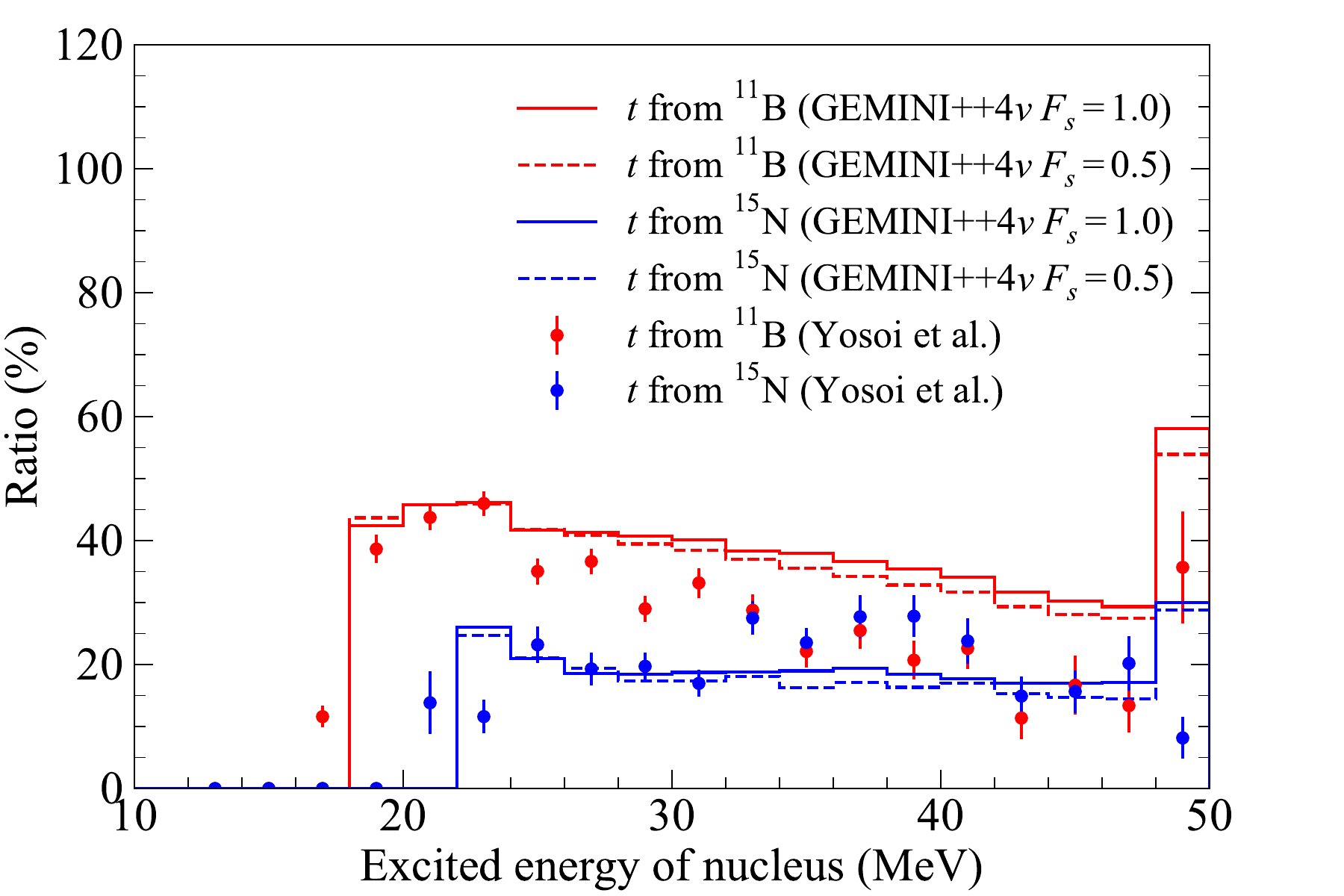}
  \includegraphics[width=0.45\textwidth]{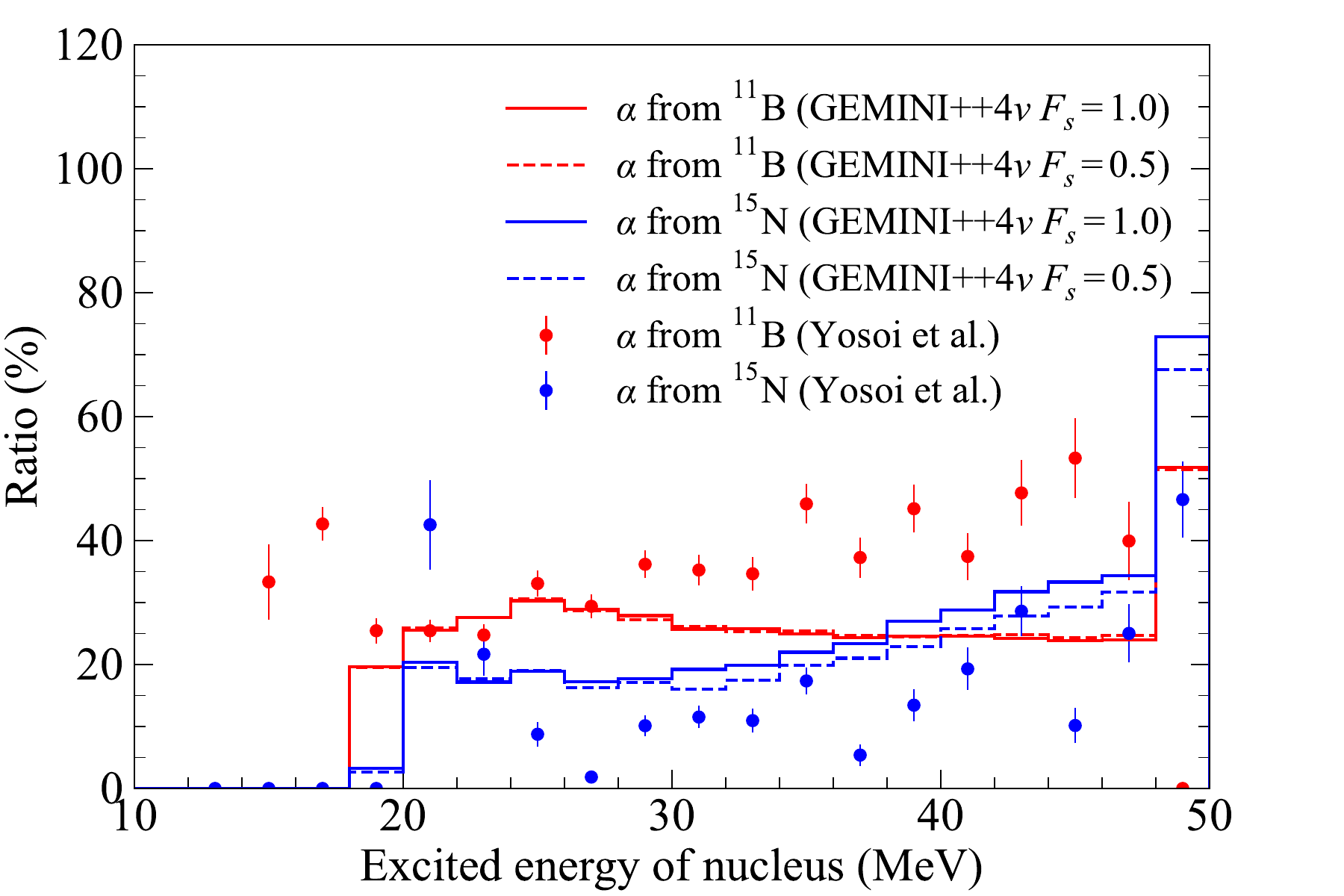}
  \caption{Comparison of measured and predicted ratios of $p$, $d$, $t$ and $\alpha$ among all four decays from $^{11}{\rm B}^*$ (red) and $^{15}{\rm N}^*$ (blue) de-excitations. The dots, solid and dashed lines denote the experimental measurements, the GEMINI++4$\nu$ predictions with the $F_s=1.0$ and $F_s=0.5$ settings, respectively.}
  \label{fig:Yosoi_Ratio_Ex}
\end{figure*}

\begin{figure}
  \centering
  \includegraphics[width=0.45\textwidth]{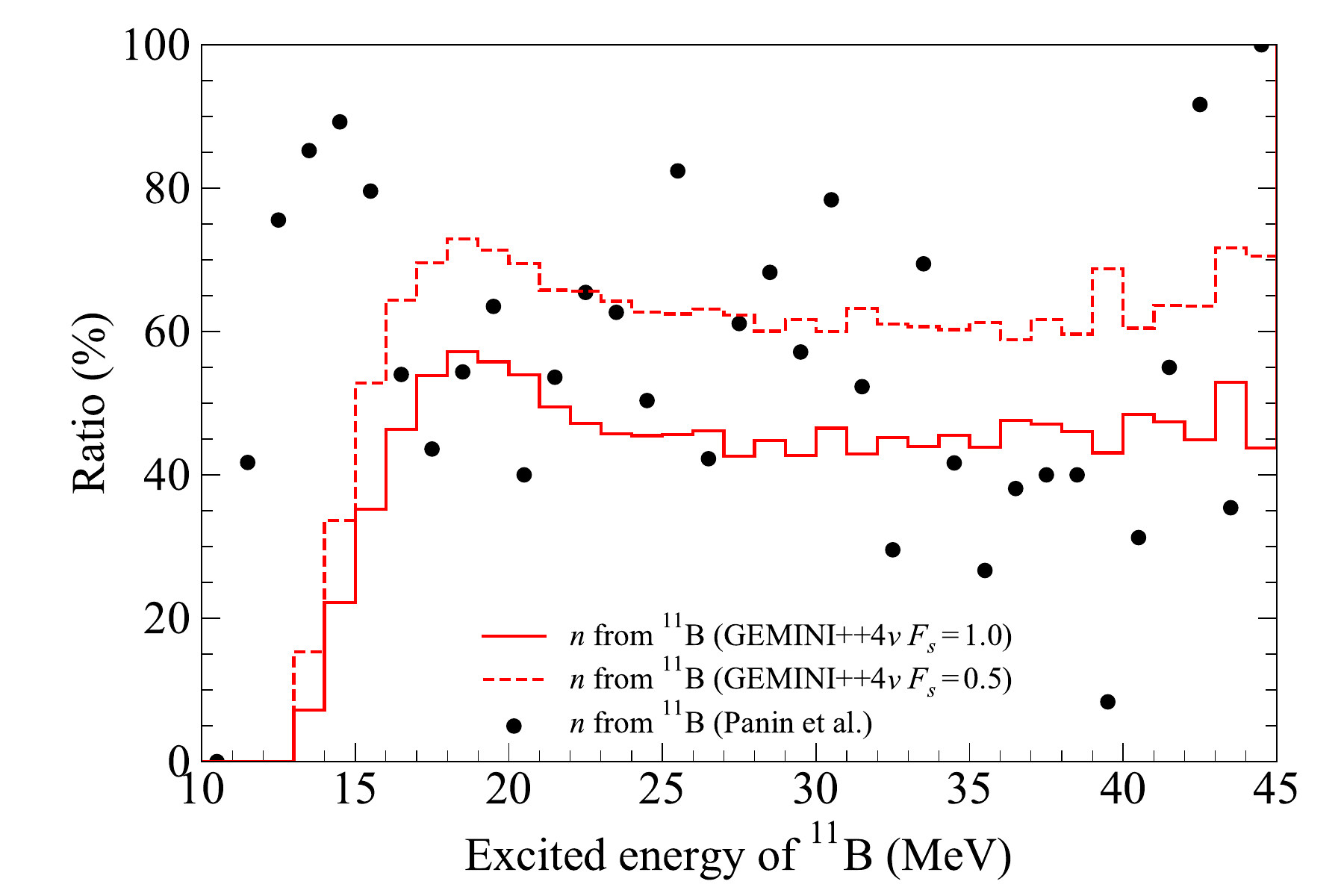}
  \caption{Comparison of measured and predicted $n$ ratios relative to all three two-body modes from $^{11}{\rm B}^*$ de-excitations. The dots, solid and dashed lines denote the experimental measurements, the GEMINI++4$\nu$ predictions with the $F_s=1.0$ and $F_s=0.5$ settings.}
  \label{fig:Panin_Shape}
\end{figure}

In the above discussions, the compared branching ratios of $^{11}{\rm B}^*$ and $^{15}{\rm N}^*$ de-excitations correspond to the excitation energy ranges of $16 ~{\rm MeV} \leq E_x \leq 35~ {\rm MeV}$ and $20 ~{\rm MeV} \leq E_x \leq 40~ {\rm MeV}$, respectively. Here we compare the ratio of each type of charged particle emission among the four types for every energy bin. The measured ratios have been plotted in Fig.~\ref{fig:Yosoi_Ratio_Ex} where only statistical error is considered. They can be determined by the digitized values from Figs. 1(b) and 2(b) in Ref.~\cite{N15}. Then we use GEMINI++4$\nu$ to calculate the corresponding ratios as shown in Fig.~\ref{fig:Yosoi_Ratio_Ex}. The differences between the $F_s=1.0$ and $F_s=0.5$ settings are relatively small. It is found that the predicted shapes are basically consistent with the measurements except for the $\alpha$ decay. The discrepancy in both $\alpha$ particles and tritons may be attributed to the cluster structure in light nuclei, such as the $\alpha$ cluster in $^{10}$Be$ $ \cite{Li:2023msp}. The $^{11}$B$^*$ nucleus can be regarded as one triton and two $\alpha$ clusters, and the decay of $^{11}$B$^* \rightarrow t + \alpha + \alpha$ is allowed in reality with a very low threshold of 11.2 MeV. The differences in the high $E_x$ region can be seen as the distinction in the kinematic energy allocation between two-body decay and three-body decay of $^{11}$B$^*$. The dominant `two-body' triton decay strength in Fig.~\ref{fig:Yosoi_Ratio} can be explained by the interactions among clusters rather than by pairing effects. In Fig.~\ref{fig:Panin_Shape}, we compare the predicted and measured ratios in every $E_x$ bin for the experiment \cite{Panin:2016div}. Both the $F_s=1.0$ and $F_s=0.5$ settings are favored, although their predictions exhibit significant discrepancies. It is difficult for us to evaluate which setting is better due to the large variation in the measured values. In Figs.~\ref{fig:Yosoi_Ratio_Ex} and~\ref{fig:Panin_Shape}, the shape comparisons shows that the GEMINI++4$\nu$ predictions do not coincidentally agree with the experimental results. 

\section{\label{sec5} Summary}

In summary, we have investigated the de-excitations of highly excited $^{11}$B$^*$ and $^{15}$N$^*$ by use of the GEMINI++ code.
Due to three issues of the critical excitation energy,
discrete levels and suppression factors, GEMINI++ can not account for the experimental data well. After addressing the three issues, we develop a code of GEMINI++4$\nu$ for neutrino experiments to handle de-excitaions of light nuclei based on the GEMINI++ code. For the settings of suppression factors, $F_s=1.0$ and $F_s=0.5$ have been assumed in this work. The predicted branching ratios of $^{11}$B$^*$ de-excitations are consistent with the measurements in the $F_s=1.0$ case. However, this setting can not explain the $^{15}$N$^*$ data. GEMINI++4$\nu$ with $F_s=0.5$ can give the better agreement with the $^{11}$B$^*$ data, and partially account for the measured branching rations of $^{15}$N$^*$ de-excitations. Therefore, the $F_s=0.5$ setting is recommended for the relevant applications. This is the first time that a statistical model code can basically reproduce both $^{11}$B$^*$ and $^{15}$N$^*$ data. Moreover, the comparisons for each energy bin show that the agreement between the GEMINI++4$\nu$ predictions and the experimental data is not coincidental. 
In the future, we plan to integrate GEMINI++4$\nu$ into widely used neutrino generators, such as GENIE and NuWro.

\acknowledgments
One of the authors (Y.J. Niu) is grateful to R.J. Charity and M. Yosoi for their help in understanding the GEMINI++ code and the experimental results, respectively. We are also grateful to Xuefeng Ding, Zhaoxiang Wu for their helpful discussions. This work is supported in part by the National Nature Science Foundation of China (NSFC) under Grants No. 12375098 and 12475136, and the Strategic Priority Research Program of the Chinese Academy of Sciences under Grant No. XDA10010100.

\end{document}